# Deep-UV bleaching of charge disorder in encapsulated graphene

Daniil Domaretskiy[1], Ned Hayward[1], Van Huy Nguyen[2], Simone Benaglia[1,2], Kornelia Indykiewicz[2], Hadrien Vignaud[1], Jing Zhang[1], Kenji Watanabe[3], Takashi Taniguchi[3], Vladimir I. Fal'ko[2], Laura Fumagalli[1], Leonid A. Ponomarenko[4], Irina V. Grigorieva[1], Andre K. Geim[1,2]

[1] Department of Physics and Astronomy, University of Manchester, Manchester, United Kingdom
[2] National Graphene Institute, University of Manchester, Manchester, United Kingdom
[3] National Institute for Materials Science, Tsukuba, Japan
[4] Department of Physics, University of Lancaster, Lancaster, United Kingdom

*Disorder masks much of the rich physics in two-dimensional electronic systems, with charged impurities often the limiting factor. In graphene, progress in reducing disorder has largely stagnated since boron nitride encapsulation was introduced a decade ago. Here we show that a brief deep-UV exposure enhances the electronic quality of encapsulated graphene – typically by two orders of magnitude - by neutralizing charged impurities within boron nitride. Following illumination, standard graphene devices exhibit numerous even-denominator fractional quantum Hall states, including non-Abelian candidates, and frequently reveal hidden superlattice minibands. Even macroscopically inhomogeneous devices, seemingly unusable for transport studies, recover after deep-UV illumination and display Landau quantization in millitesla fields. This finding provides a straightforward route to exceptional-quality graphene, enabling further exploration of interaction-driven, topological, and other quantum phenomena.*

Exploring quantum and many-body phenomena in two-dimensional (2D) electronic systems requires devices with sufficiently low disorder so that impurity scattering and charge inhomogeneity do not obscure intrinsic behavior. For decades, semiconductor-based heterostructures offered the cleanest platforms, with steady technological refinements pushing carrier mobilities ever higher (*1-3*). More recently, graphene encapsulated in hexagonal boron nitride (hBN) has emerged as a valuable alternative, delivering a two-order-of-magnitude boost in mobility and yielding electronic quality approaching that in the best GaAlAs heterostructures (*4-7*). Further improvements have been achieved by placing graphite gates within a nanometer of the graphene layer (*8*). Although this proximity screening efficiently suppresses long-range potential fluctuations, it comes at the cost of strongly weakened electron-electron interactions – the very physics that motivates the pursuit of cleaner 2D systems.

In this report, we introduce deep-UV illumination of hBN-encapsulated graphene as a tool that dramatically improves device quality by neutralizing charged impurities within hBN, thereby suppressing the associated random electrostatic potential (electron-hole puddles) and impurity scattering. The effect requires photon energies $\gtrsim 5$ eV, with lower-energy illumination proving ineffective. Deep-UV exposure for a few seconds reduces charge inhomogeneity to $\sim 10^8$ cm$^{-2}$ – a hundredfold improvement compared to unilluminated devices – and yields record mobilities up to $10^8$ cm$^2$ V$^{-1}$ s$^{-1}$, limited mainly by edge scattering in finite-size devices. The resulting quality is comparable to that previously achievable only in proximity-gated graphene and in two-terminal suspended devices (*8-10*). Critically, the technique does not suppress interaction phenomena, avoids the challenging fabrication steps required for proximity-gated or suspended devices, and is robust across diverse geometries and fabrication histories. To highlight its capabilities, we present several examples of quantum phenomena that are completely obscured by disorder in as-fabricated hBN-encapsulated graphene but become prominent following deep-UV illumination.

**Graphene before and after deep-UV**
We studied conventional Hall-bar devices made by encapsulating monolayer graphene between two hBN crystals (each 30–80 nm thick). The channel widths $W$ ranged from ~4 to 50 μm, with lengths reaching up to several times $W$. The devices (approximately 20 in total) included both newly fabricated samples and those stored for several years after prior measurements. Illumination was performed using commercial LEDs with wavelengths down to 250 nm (photon energies $E$ up to ~5.0 eV). Figure 1A compares the zero-field longitudinal resistivity $\rho_{xx}$ as a function of gate voltage $V_G$ for two representative devices before and after deep-UV exposure. Device D1 initially exhibited residual doping $n_R \approx 5\times10^{11}$ cm$^{-2}$ ($n$-type in this instance) and charge inhomogeneity $\delta n \approx 2\times10^{10}$ cm$^{-2}$. A brief exposure to 5-eV light (~10 s at liquid-helium temperature $T$) reduced the residual doping to practically zero (~$3\times10^8$ cm$^{-2}$) and suppressed charge inhomogeneity by a factor of ~200 (inset of Fig. 1A), as evidenced by the shift of the peak in $\rho_{xx}$ to zero $V_G$ and its striking narrowing (Fig. 1A). The field-effect mobility $\mu$ increased by ~100 times, reaching ~$10^8$ cm$^2$V$^{-1}$s$^{-1}$ at carrier densities $n \approx$ 3–5×10$^9$ cm$^{-2}$ and becoming limited by D1's finite $W \approx 50$ µm at higher $n$. The improvement was persistent, showing no degradation over many weeks of measurements at cryogenic temperatures, until the device was warmed to room temperature (see Supplementary Materials).

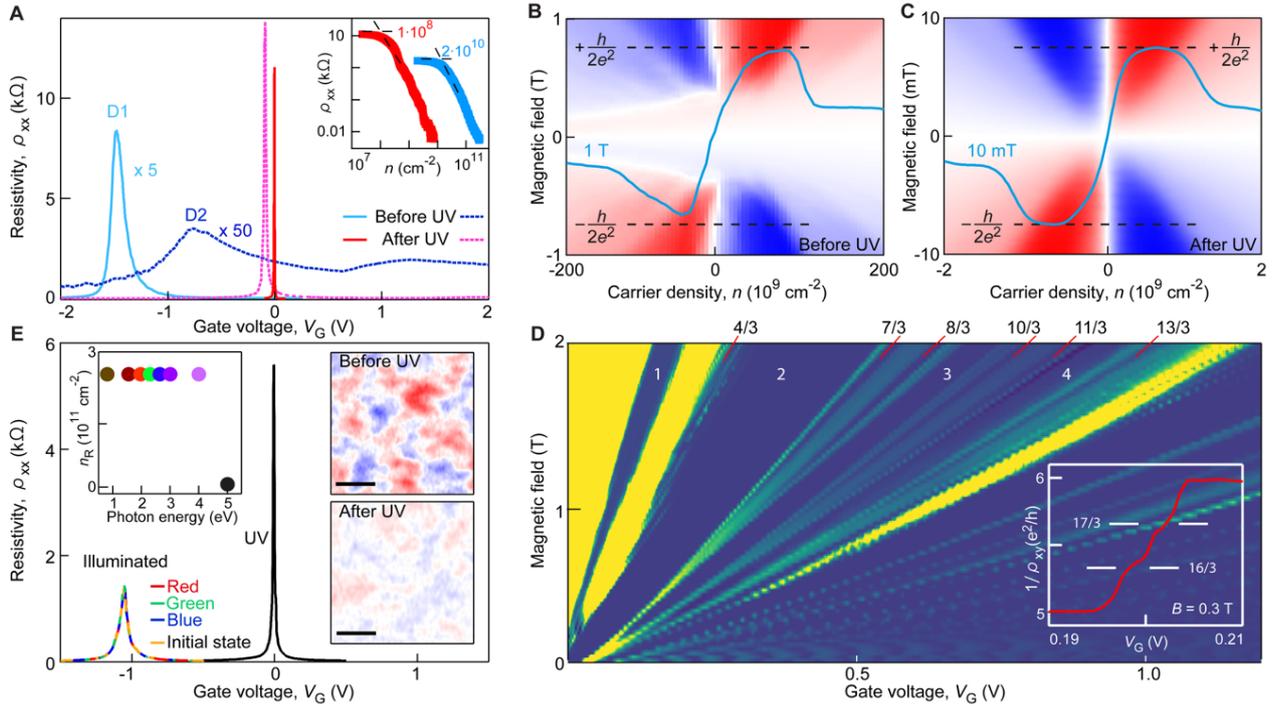

**Fig. 1| Deep-UV illumination greatly enhances graphene quality.** (**A**) Resistivity $\rho_{xx}(V_G)$ before and after illumination for devices D1 (standard homogeneity) and D2 (severe macroscopic inhomogeneity). The traces prior illuminations are scaled for clarity. Inset: Evaluation of $\delta n$ for D1 using the standard approach (see, e.g., ref. 8). After illumination, $\delta n$ decreased to ~$1\times10^8$ cm$^{-2}$ in D1 and ~$7\times10^8$ cm$^{-2}$ in D2. Data were taken at 2 K, $B = 0$. (**B** and **C**) Hall resistivity maps $\rho_{xy}(n,B)$ for D1 before (B) and after (C) illumination (blue-to-red scale: ±$h/2e^2$). Note that in (C) both $n$ and $B$ ranges are 100 times narrower compared to (B). Superimposed are $\rho_{xy}(n)$ traces in 1 T and 10 mT, respectively (their $\rho_{xy}$-scale is given by the dashed lines). (**D**) $\rho_{xx}$ map at ~50 mK for device D3 after illumination (blue-to-yellow scale: 0 to 1 kΩ). Some filling factors $\nu$ are indicated. Inset: $\rho_{xy}(V_G)$ showing FQHE plateaus at 0.3 T. (**E**) Effect of photon energy on charge homogeneity and residual doping for device D4 at 2 K (color-coded). Left inset: $n_R$ as a function of LEDs' photon energy. Illuminations were performed consecutively from infrared up to 5.0 eV, with $\rho_{xx}(V_G)$ recorded after each exposure. Right insets: electrostatic-force micrographs above a 50-nm-thick hBN crystal on graphite before (top) and after (bottom) 5-eV illumination. Both images were acquired at room temperature in the dark at a lift height of ~25 nm over the same area and use identical color scales (blue-to-red, ±50 mV). Scale bars, 1 µm. D1, D3 and D4 were graphite-gated devices with ~70-nm bottom hBN; D2 was Si-gated.

The increased homogeneity also led to profound changes in quantum transport, as shown in Fig. 1 and figs. S1-S2 in Supplementary Materials. Shubnikov-de Haas oscillations became visible in magnetic fields $B$ as low as ~3 mT (fig. S1), which yields record quantum mobilities of ~$3\times10^6$ cm$^2$ V$^{-1}$ s$^{-1}$ (*8*). Before illumination, D1 required *B* above 1 T for the quantum Hall effect (QHE) plateaus to emerge at the main filling factors $\nu = \pm2$ (Fig. 1B; fig. S2A). After deep-UV exposure, the same plateaus became fully developed in 10 mT (Fig. 1C; fig. S2B). In the two-probe geometry, using wide current contacts that did not impede edge-state transport (*8*), the QHE plateaus could be detected even in 4-5 mT (fig. S2D). This is at least an order of magnitude lower than the QHE onset reported in other 2D systems, except for proximity-gated graphene where the QHE was also observed in ~5 mT (*8*). Importantly, many-body QHE states could be resolved at record low *B*: the spin-valley degeneracy was lifted in ~0.2 T, and the fractional QHE (FQHE) emerged in 0.3 T (Fig. 1D). For comparison, the FQHE in proximity-gated graphene could be observed only above 7 T, because of weakened electron-electron interactions and despite its otherwise exceptional quality (*8*).

Even more striking were the changes induced by deep-UV exposure in macroscopically inhomogeneous devices. Before illumination, device D2 in Fig. 1A showed extreme inhomogeneity caused by an electrical breakdown between adjacent contacts. It lacked a well-defined single peak in $\rho_{xx}(V_G)$, and the resistivity curves differed between measurement geometries, rendering the device completely unusable for transport studies. Deep-UV illumination not only restored its functionality – yielding a sharply defined neutrality point (NP) with minimal residual doping (Fig. 1A) – but also resulted in transport characteristics matching those of illuminated undamaged devices. A similar recovery was observed in another highly inhomogeneous device found in this state immediately after fabrication. Importantly, every device we tested showed similarly profound improvements in quality: $\delta n$ decreased to $\lesssim 10^8$ cm$^{-2}$ in devices with graphite gates and to $\lesssim 10^9$ cm$^{-2}$ in those with Si gates.

**Photo-induced charge compensation**

In contrast to deep-UV illumination, light at longer wavelengths produced little improvement in graphene quality, as shown in Fig. 1E. Exposures from infrared to near-UV caused essentially no change in the transport characteristics of device D4, whereas 5-eV illumination markedly enhanced its quality, shifting the NP close to $V_G = 0$ and sharpening the $\rho_{xx}(V_G)$ peak. In other devices, illumination with $E < 5$ eV occasionally induced modest changes in both $\delta n$ and $n_R$, but without a clear trend – ranging from slight improvement to degradation – consistent with earlier reports (*11-14*). Electrostatic force microscopy revealed that deep-UV illumination reduced potential fluctuations above the surface of hBN crystals (insets of Fig. 1E), whereas subsequent white-light illumination restored much of the original charge disorder (fig. S3). In these figures, the electrostatic-force micrographs were obtained at room temperature, where the deep-UV-induced effect was strongly diminished (fig. S4), suggesting that the reduction of the random electrostatic potential should be much stronger at cryogenic *T*.

The observed improvements in electronic quality are tentatively attributed to an 'indirect bleaching' of electrostatic disorder via photo-induced compensation of charged impurities in hBN. Although 5-eV photons are below the hBN bandgap of ~6.0 eV, this sub-bandgap illumination can nevertheless generate mobile electrons and holes (*15*) (see Supplementary Materials). These photocarriers are expected to redistribute through the hBN bulk in response to pre-existing charge disorder, so that electrons and holes preferentially accumulate near local positive and negative potential extrema, respectively, thereby reducing long-range potential fluctuations (electron-hole puddles). Such photo-induced carrier redistribution, which tends to flatten the electrostatic landscape, is analogous to photovoltage-induced band flattening in disordered semiconductors (*16*). We speculate that, after the deep-UV light is switched off, the redistributed charges in hBN become trapped in metastable configurations, presumably involving sizable lattice relaxation (e.g., polaronic or self-trapped states) and associated high activation barriers (*15,16*). This yields a long-lived, spatially correlated trapped-charge pattern that persistently suppresses the disorder potential. The described scenario is consistent with all experimental observations, including the persistence of the low-disorder state at cryogenic temperatures and its partial survival at room temperature (Fig. 1E, figs. S3,S4). This also accounts for the observed sensitivity to subsequent illumination at longer wavelengths, which partially restores charge disorder, probably via de-trapping of the compensating charges (fig. S3). A better microscopic understanding of the compensation mechanism, which likely involves a mixture of trapping pathways and metastable defect states, would help optimize the illumination protocol for even more effective suppression of charge disorder. The Supplementary Materials provide additional information relevant to this goal, including the dependence of the achieved electronic quality on temperature, photon energy, gate type, and gate voltage.

**Hidden moiré minibands**

To illustrate how deep-UV illumination can uncover electronic features that otherwise remain hidden even in state-of-the-art graphene devices, Fig. 2 presents measurements on device D5. Before illumination, D5 exhibited transport typical of hBN-encapsulated graphene: $\rho_{xx}(n)$ peaked at the NP, reaching a few k$\Omega$, and decreased monotonically with doping over the full accessible range of $n \approx \pm 5 \times 10^{12}$ cm$^{-2}$ (note the 100-fold narrower range in Fig. 2A). The Landau fan diagrams were likewise unremarkable at both low and high fields (Figs. 2B,D). Deep-UV exposure qualitatively transformed the behavior. The NP peak became markedly sharper ($\delta n < 10^9$ cm$^{-2}$) and rose to >100 k$\Omega$, exceeding $h/e^2$ and indicating an insulating state. Landau quantization appeared at much lower $B$ (compare Figs. 2B,2D with Figs. 2C,2E), and the improved homogeneity revealed that the electron and hole Landau fans could no longer be extrapolated to the same origin (Fig. 2C), signaling a spectral gap. Finally, additional sets of Landau levels (LLs) emerged at $|n| \gtrsim 3 \times 10^{12}$ cm$^{-2}$, pointing to secondary Dirac points beyond the accessible density range (Fig. 2E).

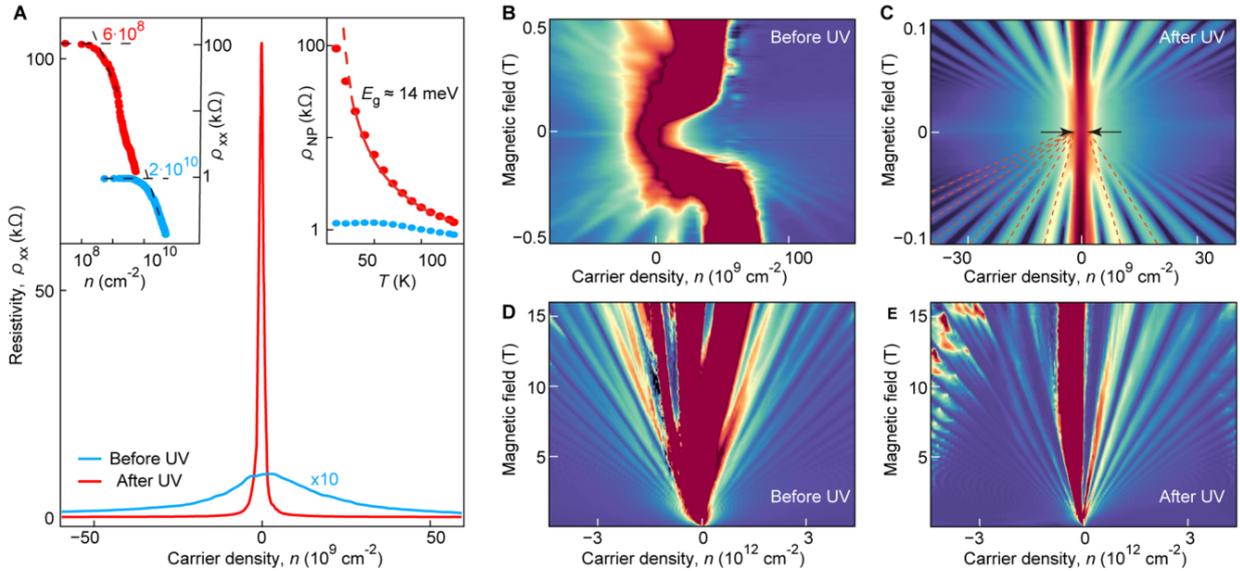

**Fig. 2 | Resolving fine electronic structure.** (**A**) $\rho_{xx}(n)$ before and after illumination (color-coded); $T \approx 2$ K, $B = 0$; device D5 with a Si gate. Left inset: Illumination reduced $\delta n$ by a factor of >30. Right inset: Temperature dependence of $\rho_{xx}$ at the NP. The red curve is a Boltzmann–Drude fit for gapped graphene, yielding $E_g \approx 14$ meV (Supplementary Materials). (**B,C**) Landau fan diagrams in low magnetic fields before (B) and after (C) illumination. (**D,E**) Same as (B,C) over a larger range of $B$. In (B, D and E), the blue-to-red scale is linear from 0 to 1 k$\Omega$. In (C), it is logarithmic ($10^2$ to $3\times10^5$ $\Omega$). Dashed lines in (C) mark maxima in $\rho_{xx}$ extrapolated to $B = 0$.

This behavior identifies D5 as a graphene-hBN moiré superlattice. From the extrapolated positions of the secondary Dirac points and the period of Brown-Zak oscillations (*17*), we inferred a twist angle of ~1.5°. The moiré-induced gap $E_g$ at the Dirac point was ~14 meV, as estimated from the temperature dependence of $\rho_{xx}$ at the NP (inset of Fig. 2A; Supplementary Materials). A similar set of hidden moiré features was also observed in two other devices (e.g., device D7 in fig. S4) that displayed no signatures of superlattice behavior prior illumination.

**Even-denominator FQHE**

The impact of deep-UV bleaching is most apparent in the clarity with which delicate many-body states emerge (Fig. 3; fig. S5). After illumination, the FQHE became visible in record-low magnetic fields (Fig. 1D) and, more strikingly, numerous even-denominator states emerged in modest fields $B \lesssim 10$ T, including several fractions not reported previously (Fig. 3; fig. S5B). At our lowest temperature of 50 mK, transport in higher orbital LLs was dominated by the re-entrant integer QHE (RIQHE) (fig. S6), a phenomenon extensively studied in semiconductor-based 2D systems and associated with many-body phases such as Wigner crystals and other charge-density-wave states that compete with FQHE liquids (*18-20*). While the RIQHE has also been reported for monolayer graphene in its $N = 2$ and $N = 3$ orbital LLs, previous observations required $B > 20$ T, making broader exploration challenging (*21,22*). Although the low-field RIQHE - made observable by deep-UV bleaching down to $B \approx 5$ T (Fig. 3D) - merits further attention, we focus here on the FQHE and therefore raised the temperature to 250 mK to suppress re-entrant features. With only residual signatures of the RIQHE persisting at this temperature (Fig. 3, fig. S6B), deep-UV-illuminated devices revealed a prominent series of half-integer FQHE states.

Figure 3A displays sharp minima in $\rho_{xx}$ and corresponding plateaus in $\rho_{xy}$ for all filling factors $\nu = k + ½$, where $k$ runs from 6 to 13 and labels graphene's spin-and-valley-polarized LLs (the first and second halves of the $\nu$ series belong to $N = 2$ and $N = 3$, respectively). Notably, half-integer states were conspicuously absent in the $N = 1$ level where numerous odd-denominator states were clearly observed (fig. S5A). Previous studies on monolayer graphene reported half-integer states only in $N = 3$, and this Landau-level specificity motivated their initial interpretation based on a parton construction (*23*). Subsequent theoretical work, however, argued that Pfaffian-like states – arising from $f$-wave pairing of composite fermions – should be energetically more favorable than parton states (*24*). Crucially, the same framework predicted an even stronger $p$-wave pairing in $N = 2$, yet half-integer FQHE had not been observed there in earlier experiments (*23*).

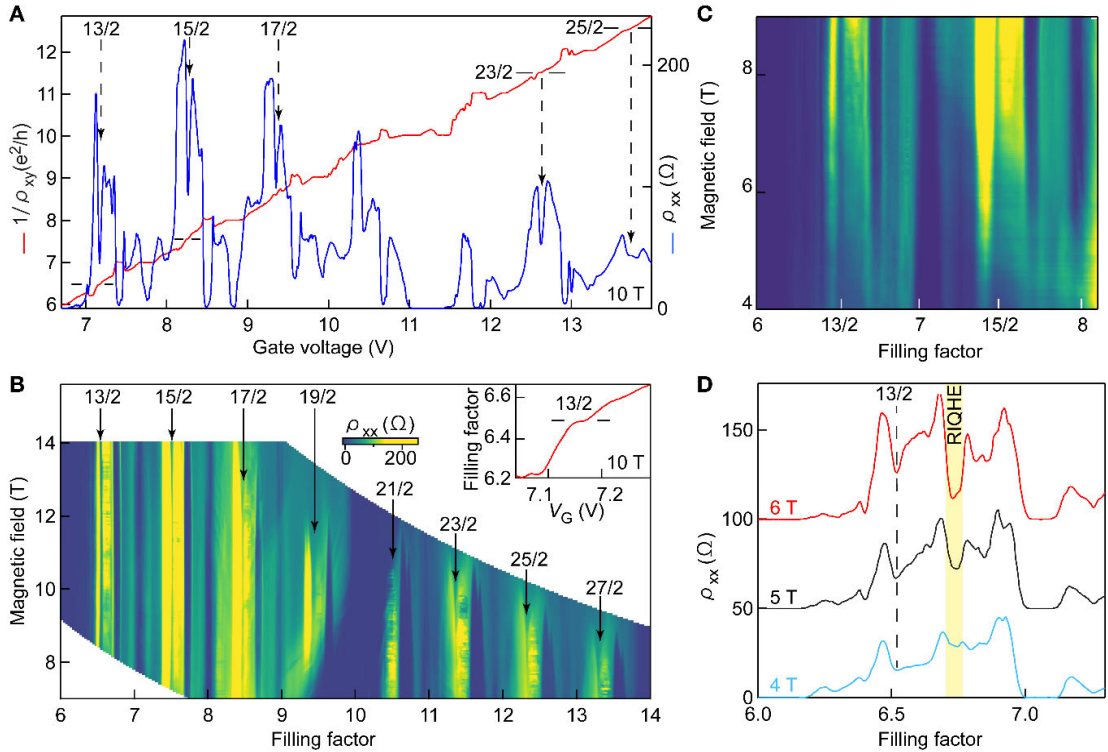

**Fig. 3| Half-integer QHE.** (**A**) $\rho_{xy}$ and $\rho_{xx}$ (red and blue curves; left and right axes, respectively). Device D3 at 10 T and 250 mK. $\rho_{xy}$ is plotted as $\nu = (h/e^2)/\rho_{xy}$. Horizontal lines mark fractional plateaus, and arrows indicate

corresponding $\rho_{xx}$ minima. (**B**) Map of $\rho_{xx}$ as a function of $B$ and the filling factor evaluated using $n$ extracted from $\rho_{xy}$ in non-quantizing fields. Arrows mark half-integer states in the second and third orbital LLs. Inset: $\rho_{xy}$ near $\nu$ = 13/2, extracted from (A). (**C**) $\rho_{xx}$ map around $\nu$ = 13/2 and 15/2 at low $B$ (same color scale as in panel B). (**D**) Representative traces from (C) showing persistence of the 13/2 state down to $B \approx 5$ T.

In contrast to the previous report but consistent with these theoretical expectations, our deep-UV-illuminated devices exhibit their most robust half-integer sequence in $N$ = 2, with weaker counterparts in $N$ = 3 (Fig. 3). Together with the absence of half-integer states in $N$ = 1, this behavior strongly supports an interpretation of the observed half-integer FQHE in terms of BCS-like pairing of composite fermions (*24*) and helps resolve the existing controversy. Experimentally, the $\nu = k + 1/2$ states in $N$ = 2 are exceptionally robust, with $\nu$ = 13/2 persisting down to 5 T even at an elevated temperature of 250 mK (Figs. 3C,D). These states are direct analogues of the celebrated $\nu$ = 5/2 state originally discovered in GaAlAs heterostructures (*25,26*) and are expected to host Moore-Read quasiparticles, which attract sustained interest due to their non-Abelian statistics and potential utility for topologically protected quantum computation (*27*).

Another even-denominator FQHE state that has neither been reported previously nor predicted for graphene, yet emerged prominently after deep-UV bleaching is $\nu$ = 2 + 3/10 (fig. S5). Because the $N$ = 1 level was free of re-entrant QHE behavior, we were able to probe this fractional state down to 50 mK, where it exhibited both vanishing $\rho_{xx}$ and a well-defined Hall plateau (fig. S5B). The activation energy was evaluated as ~1.0 K (inset of fig. S5B). The observed fraction is analogous to $\nu$ = 3/10 in semiconductor-based 2D systems and is usually interpreted as BCS-like pairing of composite fermions involving four flux quanta per electron (*28*). This pairing is predicted to form anti-Pfaffian quasiparticles, which are also non-Abelian (*28,29*).

**Outlook**

Deep-UV illumination offers a simple and reliable way to achieve exceptional-quality graphene, opening access to phenomena previously obscured by charge disorder. The resulting homogeneity is essential for studies of fragile many-body phases, including even-denominator and non-Abelian fractional states, symmetry-breaking phases, unconventional superconductivity, quantum magnetism, and excitonic condensates. Because the method targets charged impurities in the encapsulating hBN without altering the graphene itself, it should extend naturally to bilayers, multilayers, magic-angle and other twisted graphene systems, and potentially to a broad range of hBN-encapsulated 2D materials. This development should make deep-UV illumination as routine as hBN encapsulation, broadening access to interaction-driven and topological phenomena across van der Waals heterostructures.